\documentclass[twocolumn,pra,aps,showpacs]{revtex4}

\usepackage{mathptmx}
\usepackage{subfigure}
\usepackage{psfrag,graphicx}
\usepackage{dcolumn}
\usepackage{amsmath,amssymb}
\usepackage{bm}
\usepackage{color}
\usepackage{latexsym}
\usepackage{epstopdf}
\usepackage{color}
\usepackage[english]{babel}
\usepackage{latexsym}
\usepackage{psfrag,graphicx}
\usepackage{subfigure}
\usepackage{amsmath}
\usepackage{amssymb}
\usepackage{amsfonts}
\usepackage{bm}
\usepackage{natbib}
\usepackage{epstopdf}
\usepackage{appendix}

\definecolor{mygrey}{gray}{0.35}
\definecolor{myblue}{rgb}{0.2,0.2,0.8}
\definecolor{myzard}{cmyk}{0,0,0.05,0}
\definecolor{mywhite}{rgb}{1,1,1}
\definecolor{mywhite}{rgb}{1,1,1}
\definecolor{myred}{rgb}{1,0.,0.3}

\usepackage[colorlinks=true,citecolor=myblue,linkcolor=myred]{hyperref}

\def\ba{\begin{align}}
\def\enda{\end{align}}
\def\bi{\begin{itemize}}
\def\ei{\end{itemize}}

\def\be{\begin{equation}}
\def\ee{\end{equation}}
\def\bea{\begin{eqnarray}}
\def\eea{\end{eqnarray}}
\def\bse{\begin{subequations}}
\def\ese{\end{subequations}}

\begin{document}
\title{Force sensors with precision beyond the standard quantum limit}
\author{Peter A. Ivanov}
\affiliation{Department of Physics, St. Kliment Ohridski University of Sofia, James Bourchier 5 blvd, 1164 Sofia, Bulgaria}

\begin{abstract}
We propose force sensing protocols using linear ion chain which can operate beyond the quantum standard limit. We show that oscillating forces that are off-resonance with the motional trap frequency can be detected very efficiently by using quantum probes represented by various spin-boson models. We demonstrate that the temporal evolution of a quantum probe described by the Dicke model can be mapped on the nonlinear Ramsey interferometry which allows to detect far-detuned forces simply by measuring the collective spin populations. Moreover, we show that the measurement uncertainty can reach the Heisenberg limit by using initial spin correlated states, instead of motional entangled states. An important advantage of the sensing technique is its natural robustness against the thermally induced dephasing, which extends the coherence time of the measurement protocol. Furthermore, we introduce sensing scheme that utilize the strong spin-phonon coupling to improve the force estimation. We show that for quantum probe represented by quantum Rabi model the force sensitivity can overcome those using simple harmonic oscillator as a force sensor.
\end{abstract}

\maketitle

\section{Introduction}
A precise measurement of very weak forces using nano-mechanical oscillators has broad and important applications ranging from atomic force microscopy \cite{Mamin2001,Rugar2004} to testing the fundamental physics \cite{Geraci2008,Arvanitaki2013}. Various quantum-optical systems can be used to detect very weak forces with sensitivity below the attonewton range including for example nanomechanical oscillator coupled to a microwave cavity \cite{Teufel2009}, carbon nanotubes \cite{Moser2013} and trapped ions \cite{Biercuk2010,Knunz2010,Gloger2015,Shaniv2016}. In particular, trapped ions are promising quantum system with application in the weak force sensing due to the broad tunability of the trapping frequencies as well as the high-precision read-out of the spin and vibrational states. As has been shown in \cite{Maiwald2009} a force sensitivity in the range of 1 yN ($10^{-24}$ N) per $\sqrt{\rm{Hz}}$ can be achieved when the external driving force oscillates exactly in resonance with the ion's motional frequency. Another approach considers force sensing protocols capable to detect forces that are off-resonance with the trapping frequency by using various spin-boson models \cite{Ivanov2015,Ivanov2016}. 
Recently, the detection of weak low-frequency forces with sensitivity as low as 0.5 aN ($10^{-18}$ N) per $\sqrt{\rm{Hz}}$ was experimentally demonstrated using Doppler velocimetry technique \cite{Shaniv2016}.

In this work we study the temporal evolution of the collective spin-boson Jahn-Teller model and show that it can be mapped on nonlinear Ramsey interferometer for measuring very weak forces. We consider force sensor protocols that utilize the laser induced coupling between the collective spin states and single vibrational mode and show that they can be used to detect very efficiently weak forces that are off-resonance with the ion's trap frequency. We demonstrate that low-frequency forces that are below the trapping frequency can be detected by using probe represented by the Dicke model \cite{Dicke1954}. We show that for force detuning much higher than the spin-phonon coupling the relevant force information is mapped into the collective spin-degree of freedom. This allows to use the spin correlation instead of motional entangled states \cite{Munro2002} to improve the force sensitivity. We show that for initial uncorrelated spin states, the force sensitivity is short noise limited, while for the initial maximum entangled spin state the force sensitivity is Heisenberg limited. The main advantage of the proposed sensing protocol is its natural robustness against the thermally induced spin dephasing, which avoids the applications of additional dynamical decoupling techniques during the force estimation. The absence of residual spin-vibrational interaction extends the coherence time of the sensing protocol and allows to use ion chain which is not laser cooled to the vibrational ground state.

Furthermore, we introduce force sensor technique which is capable to detect time-varying forces with detuning smaller than the spin-phonon coupling by mapping the relevant force information into the vibrational degree of freedom. Here the quantum probe is represented by the quantum Rabi model describing the dipolar interaction between the single vibrational mode and effective spin states \cite{Rabi1936}. We show that the force sensitivity of our technique can overcome the sensitivity which is achieved by using simple harmonic oscillator as a force sensor. Moreover, thanks to the strong spin-phonon coupling our sensitivity can overcome even the best sensitivity that can be achieved when the force oscillates at resonance with the ion's trap frequency.

The paper is organized as follows: In Sec. \ref{JTmodel} we introduce the Janh-Teller spin-boson model which we use as a quantum probe sensitive to very weak forces. In Sec. \ref{Implementation} we discuss the physical implementation of the model using linear ion crystal. In Sec. \ref{weakregime} we introduce nonlinear Ramsey type sensing protocol capable to detect far-detuned forces by measuring the collective spin population. Here the quantum probe sensitive only to one force component is represented by the Dicke model. For detuning much higher than the spin-phonon coupling the model is mapped on the one-axis twisting model. We show that the technique is not-sensitive to thermally induced dephasing. It is shown that using the initial spin correlation states can improve the force sensitivity to the Heisenberg limit. In Sec. \ref{strongregime} we consider sensing protocol of time-varying forces with detuning smaller than the spin-boson coupling. Thanks to this we show that the minimal detectable force of our protocol overcomes the best sensitivity which can be achieved using simple harmonic oscillator as a force sensor. Finally, in Sec. \ref{conclusions} we summarize our findings.

\section{The model}\label{JTmodel}

We consider a model in which an ensemble of $N$ two-state atoms interact with two boson modes via Jahn-Teller coupling \cite{Larson2008}
\begin{eqnarray}
&&\hat{H}=\hat{H}_{0}+\hat{H}_{\rm JT}+\hat{H}_{F},\notag\\
&&\hat{H}_{0}=\hat{H}_{\rm b}+\hat{H}_{\rm s}=\hbar\delta_{x}\hat{a}_{x}^{\dag}\hat{a}_{x}+\hbar\delta_{y}\hat{a}_{y}^{\dag}\hat{a}_{y}+\hbar\Delta \hat{J}_{z},\notag\\
&&\hat{H}_{\rm JT}=\frac{2\hbar g_{x}}{\sqrt{N}}\hat{J}_{x}(\hat{a}_{x}^{\dag}+\hat{a}_{x})
+\frac{2\hbar g_{y}}{\sqrt{N}}\hat{J}_{y}(\hat{a}_{y}^{\dag}+\hat{a}_{y}),\notag\\
&&\hat{H}_{F}=\sqrt{N}F_{x}(\hat{a}_{x}^{\dag}+\hat{a}_{x})+\sqrt{N}F_{y}(\hat{a}_{y}^{\dag}+\hat{a}_{y}).\label{model}
\end{eqnarray}
Here $\hat{H}_{0}$ contains the free boson term where $a_{\alpha}^{\dag}$, $a_{\alpha}$ ($\alpha=x,y$) are the creation and annihilation operators corresponding to oscillator with frequency $\omega_{\alpha}$. The term $\hat{H}_{\rm s}$ describes the interaction between the collection of spins and the external applied magnetic field with strength $\Delta$. The second term in (\ref{model}) is the Jahn-Teller spin-boson interaction with coupling strength $g_{\alpha}$, where $\hat{J}_{\beta}=\frac{1}{2}\sum_{k=1}^{N}\sigma_{k}^{\beta}$ ($\beta=x,y,z$) are the collective spin operators with $\sigma_{k}^{\beta}$ being the Pauli operator for spin $k$. The last term in (\ref{model}) describes the action of force which displaces the two bosonic modes with strength $F_{\alpha}$.

The collective spin basis consists of the set of the eigenvectors $\{\left|j,m\right\rangle\}$ of the two commuting operators $\hat{J}^{2}\left|j,m\right\rangle=j(j+1)\left|j,m\right\rangle$ and $\hat{J}_{z}\left|j,m\right\rangle=m\left|j,m\right\rangle$ ($m=-j,\cdots,j$) with $j=\frac{N}{2}$. The total Hilbert space is spanned in the basis $\{\left|j,m\right\rangle\otimes\left|n_{x},n_{y}\right\rangle\}$, where $\left|n_{\alpha}\right\rangle$ is the Fock state of the bosonic mode with occupation number $n_{\alpha}$.

For general non-equal couplings $g_{x}\neq g_{y}$ and $F_{x}=F_{y}=0$ the model possesses a discrete $\mathbb{Z}$ symmetry. Indeed, the parity operator defined by
\begin{equation}
\hat{\Pi}=\hat{\Pi}_{\rm s}\otimes\hat{\Pi}_{\rm b},\quad \hat{\Pi}_{\rm s}=\sigma_{1}^{z}\otimes\cdots\otimes\sigma_{N}^{z},\quad \hat{\Pi}_{\rm b}=(-1)^{\hat{a}_{x}^{\dag}\hat{a}_{x}+\hat{a}_{y}^{\dag}\hat{a}_{y}},\label{parity}
\end{equation}
transform $\hat{J}_{x,y}\rightarrow-\hat{J}_{x,y}$ and $\hat{a}_{x,y}\rightarrow-\hat{a}_{x,y}$ which implies that $\hat{H}\rightarrow\hat{H}$.

Let us discuss a few well known limits of our model. First, when either ($g_{y}=0$, $g_{x}\neq0$) or ($g_{x}=0$, $g_{y}\neq0$) the model is equivalent to the Dicke model \cite{Dicke1954}. For the particular case of single spin $j=1/2$ it reduces to the quantum Rabi model \cite{Rabi1936}. In the symmetric case $\delta_{x}=\delta_{y}$ and $g_{x}=g_{y}$ the model (\ref{model}) describes the U(1) invariant Jahn-Teller spin-boson interaction. In the limit of $N=1$ the model reduces to $E\otimes e$ symmetrical Jahn-Teller model which has been shown to possesses an effective gauge potential description \cite{Larson2009}. On the other hand in the semiclassical limit $N\gg 1$ the model exhibits a magnetic structural phase transition \cite{Ivanov2013}.

In the following, we discuss the physical realization of the Hamiltonian (\ref{model}) using linear ion crystal.

\section{Physical Implementation}\label{Implementation}
Trapped ions are a suitable system to implement the Jahn-Teller spin-boson coupling in a two-level system by driving simultaneously red- and blue-sideband transitions with external laser field \cite{Wineland1998,Schneider2012}. Consider a linear ion crystal of $N$ trapped ions with mass $m$ confined in a linear Paul trap along the $z$ axis with trap frequencies $\omega_{\beta}$. The position operator of ion $k$ is given by
\begin{equation}
\hat{\vec{r}}_{k}=\delta r_{x,k}\vec{e}_{x}+\delta \hat{r}_{y,k}\vec{e}_{y}+(z_{k}^{0}+\delta \hat{r}_{z,k})\hat{e}_{z},
\end{equation}
where $z_{k}^{0}$ are the equilibrium positions along the trapping $z$ axis and $\delta \hat{r}_{\beta,k}$ are the displacement operators around the equilibrium positions. In terms of collective modes the latter can be written as $\delta\hat{r}_{\beta,k}=\sum_{p=1}^{N}b_{k,p}^{\beta}\sqrt{\frac{\hbar}{2m\omega_{p,\beta}}}(\hat{a}_{p,\beta}^{\dag}+\hat{a}_{p,\beta})$ where $\hat{a}_{p,\beta}^{\dag}$ and $\hat{a}_{p,\beta}$ are respectively the creation and annihilation operators of the $p$th vibration mode along $\beta$ direction with corresponding vibrational frequency $\omega_{p,\beta}$ and $b_{k,p}^{\beta}$ are the normal mode wave functions \cite{James1998,Steane1997}.
We assume that the two-level system of each ion consists of two metastable levels. Here we consider an atomic $\Lambda$-type system where the Jahn-Teller coupling is driven by a Raman-type interaction. For example such a level structure occurs in the hyperfine levels of $^{171}$Yb$^{+}$ ion where the qubit states are formed by the magnetic insensitive states $\left|\uparrow\right\rangle=|F=1,m_{F}=0\rangle$ and $\left|\downarrow\right\rangle=|F=0,m_{F}=0\rangle$ with transition frequency $\omega_{0}$ \cite{Olmschenk2007}. The interaction-free Hamiltonian describing the ion crystal is given by
\begin{equation}
\hat{H}_{\rm free}=\hbar\omega_{0}\hat{J}_{z}+\hbar\sum_{p=1}^{N}\sum_{\beta=x,y,z}\omega_{p,\beta}\hat{a}_{p,\beta}^{\dag}\hat{a}_{p,\beta}.
\end{equation}
Consider that the linear ion crystal is simultaneously addressed by bichromatic laser fields in a Raman configuration along two transverse orthogonal $x$ and $y$ directions with laser frequencies beat notes $\omega_{r,\alpha}=\omega_{0}-\Delta-(\omega_{1,\alpha}-\delta_{\alpha})$ and $\omega_{b,\alpha}=\omega_{0}-\Delta+(\omega_{1,\alpha}-\delta_{\alpha})$ which induce a transition between the qubit states via an auxiliary excited state. Here $\delta_{x}$ and $\delta_{y}$ are the detunings to the center-of-mass vibrational modes along the two transverse directions so that $\omega_{1,x}=\omega_{x}$ and $\omega_{1,y}=\omega_{y}$, while $\Delta$ is the detuning of the AC-Stark shifted states with respect to $\omega_{0}$. The Hamiltonian describing laser-ion interaction, after making optical rotating-wave approximation, is given by \cite{Lee2005}
\begin{eqnarray}
\hat{H}_{I}&=&\hbar\Omega_{x}\sum_{k=1}^{N}\{\sigma_{k}^{+}e^{ik_{x}\delta \hat{r}_{x,j}-i\phi_{x}}(e^{-i\omega_{r,x}t}+e^{-i\omega_{b,x}t})+{\rm h.c.}\}\notag\\
&&+\hbar\Omega_{y}\sum_{k=1}^{N}\{\sigma_{k}^{+}e^{ik_{y}\delta \hat{r}_{y,j}-i\phi_{y}}(e^{-i\omega_{r,y}t}+e^{-i\omega_{b,y}t})\notag\\
&&+{\rm h.c.}\}.\label{HI}
\end{eqnarray}
Here $\Omega_{\alpha}$ are the two-photon Rabi frequencies, $\vec{k}_{\alpha}$ are the laser wave vectors ($k_{\alpha}=|\vec{k}_{\alpha}|$) and $\phi_{\alpha}$ are the respective laser phases which we set to $\phi_{x}=\pi/2$ and $\phi_{y}=0$. Next we assume the Lamb-Dicke limit and transform the Hamiltonian (\ref{HI}) in the rotating-frame with respect to $\hat{U}_{R}(t)=e^{-i(\omega_{0}-\Delta) t \hat{J}_{z}-i\sum_{\alpha}\sum_{p=1}^{N}(\omega_{p,\alpha}-\delta_{\alpha})t\hat{a}^{\dag}_{\alpha}\hat{a}_{\alpha}}$ which yields
\begin{equation}
\hat{H}_{0}+\hat{H}_{\rm JT}=\hat{U}_{R}^{\dag}(\hat{H}_{\rm free}+\hat{H}_{I})\hat{U}_{R}-i\hbar\hat{U}_{R}^{\dag}\partial_{t}\hat{U}_{R},\label{H0HI}
\end{equation}
where the spin-phonon couplings are $g_{\alpha}=\eta_{\alpha}\Omega_{\alpha}$ with $\eta_{\alpha}=k_{\alpha}r_{0,\alpha}$ stand for the Lamb-Dicke parameters ($\eta_{\alpha}\ll1$) with $r_{0,\alpha}=\sqrt{\hbar/2m\omega_{\alpha}}$ being the spread of the oscillator center-of-mass ground-state wave function. In Eq. (\ref{H0HI}) we have assumed motional rotating-wave approximation which is fulfilled as long as $|\omega_{1,\alpha}-\omega_{p\neq 1,\alpha}|\gg g_{\alpha},|\delta_{\alpha}|$. The latter condition ensures that all vibrational modes can be neglected except the center-of-mass mode.

The last symmetry breaking term in (\ref{model}) represents the action of the external driving force that displaces a vibrational amplitude of the transverse center-of-mass vibrational modes. Indeed, the action of the force is described by $\hat{H}_{F}(t)=\sum_{\alpha}\sum_{k=1}^{N}F_{d,\alpha}(t)\delta \hat{r}_{\alpha,k}$, where we assume $F_{d,\alpha}(t)=f_{d,\alpha}\cos[(\omega_{\alpha}-\delta_{\alpha})t]$ with $f_{d,\alpha}$ being the amplitude of the force. By transforming $\hat{H}_{F}(t)$ to the rotating frame by means of $\hat{U}_{R}(t)$ and neglecting the fast-rotating terms we obtain $\hat{H}_{F}$ where $F_{\alpha}=r_{0,\alpha}f_{d,\alpha}/2$. In the following we introduce sensing protocols that are capable to detect the force amplitude $f_{d,\alpha}$ by observing the time-evolution either of the collective spin populations or the mean phonon number.

Finally, we note that the unitary operator $\hat{U}_{R}(t)$ commutes with the observable of interest such as $\hat{J}_{z}$ and $\hat{a}_{\alpha}^{\dag}\hat{a}_{\alpha}$ which implies that this transformation does not introduce additional error during the force estimation.
\section{Sensing low-frequency forces}\label{weakregime}
\begin{figure}
\includegraphics[width=0.45\textwidth]{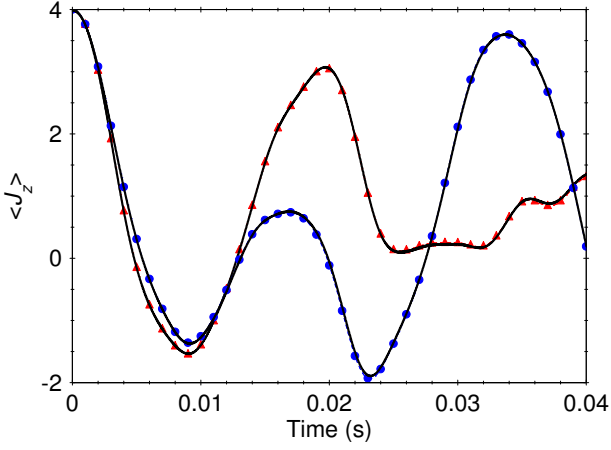}
\caption{(Color online) Time-evolution of the expectation value of $\hat{J}_{z}$ operator for a system of $N=8$ spins.
We assume an initial state $|\Psi(0)\rangle=|j,j\rangle|0_{x},0_{y}\rangle$. We compare the numerical solution of the time-dependent Schr\"odinger equation with Hamiltonian (\ref{model}) (solid lines) with the solution using effective Hamiltonian (\ref{Heff}) for $\Delta=\chi_{x}\chi_{y}$ (blue circles) and $\Delta=2\chi_{x}\chi_{y}$ (red triangles). The parameters are set to $g_{x}=5$ kHz, $g_{y}=3$ kHz, $\delta_{x}=-85$ kHz, $\delta_{y}=-80$ kHz, $f_{d,x}=10$ yN and $f_{d,y}=15$ yN.}
\label{fig1}
\end{figure}
We begin by considering the weak coupling regime of our model (\ref{model}) in which the detuning $\delta_{\alpha}$ of the driving force is much higher than the spin-phonon coupling $g_{\alpha}$ ($|\delta_{\alpha}|\gg g_{\alpha}$). In that case the center-of-mass modes are only virtually excited, thereby they can be adiabatically eliminated from the dynamics. This can be carried out by applying the canonical transformation $\hat{U}=e^{\hat{S}}$ to the Hamiltonian (\ref{model}) such that $\hat{H}_{\rm eff}=\hat{U}^{\dag}\hat{H}\hat{U}$, where the anti-Hermitian operator $\hat{S}$ is given by
\begin{eqnarray}
\hat{S}&=&\frac{2g_{x}}{\delta_{x}\sqrt{N}}\hat{J}_{x}(\hat{a}_{x}-\hat{a}_{x}^{\dag})
+\frac{2g_{y}}{\delta_{y}\sqrt{N}}\hat{J}_{y}(\hat{a}_{y}-\hat{a}_{y}^{\dag})\notag\\
&&+\sqrt{N}\frac{F_{x}}{\delta_{x}}(\hat{a}_{x}-\hat{a}_{x}^{\dag})+\sqrt{N}\frac{F_{y}}{\delta_{y}}(\hat{a}_{y}-\hat{a}_{y}^{\dag}).
\end{eqnarray}
Keeping only the leading terms of order of $g_{\alpha}/\delta_{\alpha}$ the effective Hamiltonian becomes $\hat{H}_{\rm eff}=\hat{H}_{0}+\frac{1}{2}[\hat{H}_{\rm JT}+\hat{H}_{F},\hat{S}]+\hat{H}^{\prime}$ which yields
\begin{eqnarray}
&&\hat{H}_{\rm eff}=\hat{H}_{\rm_{spin}}+\hat{H}_{\rm{res}}+\hat{H}^{\prime},\notag\\
&&\hat{H}_{\rm_{spin}}=\hbar\Delta \hat{J}_{z}-\frac{4\hbar g_{x}^{2}}{N\delta_{x}}\hat{J}_{x}^{2}-\frac{4\hbar g_{y}^{2}}{N\delta_{y}}\hat{J}_{y}^{2}-\frac{4 g_{x}F_{x}}{\delta_{x}}\hat{J}_{x}
-\frac{4g_{y}F_{y}}{\delta_{y}}\hat{J}_{y},\notag\\
&&\hat{H}_{\rm_{res}}=\hat{H}_{\rm b}+\frac{2i\hbar g_{x}g_{y}}{N\delta_{x}\delta_{y}}\hat{J}_{z}\{(\delta_{x}+\delta_{y})(\hat{a}_{x}^{\dag}\hat{a}_{y}-{\rm h.c.})\notag\\
&&-(\delta_{x}-\delta_{y})(\hat{a}_{x}^{\dag}\hat{a}_{y}^{\dag}-{\rm h.c.})\},\label{Heff}
\end{eqnarray}
where the we have omitted the constant terms. The result indicates that the phonon degree of freedom mediates an effective spin-spin interaction described by the nonlinear quadratic collective spin operators in $\hat{H}_{\rm spin}$. In addition to it the effect of the symmetry-breaking term $\hat{H}_{F}$ is to induce transition between the individual spin states which are captured by the last two linear collective spin operators in $\hat{H}_{\rm spin}$. The term $\hat{H}_{\rm res}$ is the residual spin-phonon interaction, which does not couple spins at different sites, but rather describes processes in which phonon excitations are created and respectively reabsorbed by the same spin. Note that as long as the quantum oscillators are in their ground states the term $\hat{H}_{\rm res}$ does not affect the collective spin dynamics and thus it can be neglected. Finally, the term $\hat{H}^{\prime}=\frac{1}{3}[[\hat{H}_{\rm JT}+\hat{H}_{F},\hat{S}],\hat{S}]+\ldots$ contains high-order terms in the spin-phonon interaction which we neglected as long as $|\delta_{\alpha}|\gg g_{\alpha}$.
\begin{figure}
\includegraphics[width=0.45\textwidth]{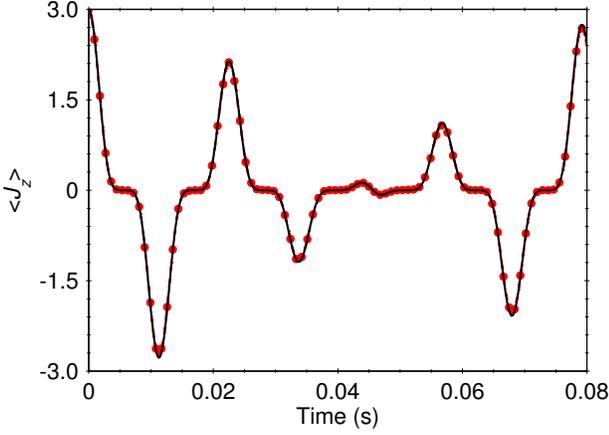}
\caption{(Color online) Time-evolution of the expectation value of $\hat{J}_{z}$ operator for a system of $N=6$ spins.
We assume an initial thermal vibrational state with average phonon number $\bar{n}=0.6$. We compare the numerical solution of the time-dependent Schr\"odinger equation with Hamiltonian (\ref{model}) (solid lines) with the solution using effective Hamiltonian (\ref{OATM}) (red circles). The parameters are set to $g_{x}=5$ kHz, $\delta_{x}=60$ kHz, $r_{0,x}=14.5$ nm, $f_{d,x}=1.5$ yN.}
\label{fig2}
\end{figure}

Hence, in the weak coupling regime the model (\ref{model}) is mapped into the generalized Lipkin-Meshkov-Glick (LMG) Hamiltonian \cite{Lipkin1965}. As can be seen from Eq. (\ref{Heff}) the sign of the coupling strengths of the non-linear spin terms depend on the sign of the detunings $\delta_{\alpha}$, thus one could achieve ferromagnetic interaction $\delta_{\alpha}>0$ or respectively anti-ferromagnetic interaction $\delta_{\alpha}<0$. It is important to note that following the same line as in \cite{Unanyan2003,Unanyan2005} our effective Hamiltonian (\ref{Heff}) in the anti-ferromagnetic regime possesses supersymmetric structure at the special point $\Delta=\chi_{x}\chi_{y}$ where we define $\chi_{\alpha}^{2}=4g_{\alpha}^{2}/(N|\delta_{\alpha}|)$. Indeed, it is straightforward to show that at this point the Hamiltonian (\ref{Heff}) takes the form
\begin{equation}
\hat{H}_{\rm eff}=\hbar(\chi_{x} \hat{J}_{x}+i\chi_{y}\hat{J}_{y}+\gamma)(\chi_{x} \hat{J}_{x}-i\chi_{y}\hat{J}_{y}+\gamma^{*})-\hbar|\gamma|^{2},
\end{equation}
where $\gamma=\mu_{x}\chi_{x}+i\mu_{y}\chi_{y}$ and $\mu_{\alpha}=F_{\alpha}N/2g_{\alpha}$. Figure (\ref{fig1}) shows the time-evolution of the expectation value of $\hat{J}_{z}$ according the model (\ref{model}) compared with the effective Hamiltonian (\ref{Heff}). As expected, the effective picture based on LMG model is very accurate in the weak coupling regime.

Let us now focus on the sensing protocol capable to detect only one of the force components, namely $f_{d,x}$. Thus in the following we set $\Delta=0$, $g_{y}=0$ such that the quantum probe is represented by the Dicke Hamiltonian describing the dipolar interaction between the ensemble of $N$ atoms with the single vibrational mode. In the limit $|\delta_{x}|\gg g_{x}$ the effective Hamiltonian reduces to the one-axis twisting Hamiltonian
\begin{equation}
\hat{H}_{\rm eff}=-\hbar\chi^{2} \hat{J}_{x}^{2}-\hbar\Omega_{f} \hat{J}_{x},\label{OATM}
\end{equation}
where we define $\chi=\chi_{x}$ and $\Omega_{f}=2g_{x}r_{0,x}f_{d,x}/\hbar\delta_{x}$. In that case as can be seen from Eq. (\ref{Heff}) the residual spin-phonon term $\hat{H}_{\rm res}$ vanishes automatically. Moreover, it is straightforward to show that even the high-order terms in the residual spin-phonon coupling vanishes such that we have $\hat{H}^{\prime}=0$ which indicates that for $|\delta_{x}|\gg g_{x}$ the model is exactly mapped into the one-axis twisting Hamiltonian (\ref{OATM}).

The nonlinear Hamiltonian (\ref{OATM}) has been proposed for practical applications to quantum metrology, since it can produce squeezed-spin states \cite{Ma2011}. For example, such interaction is used to perform precision measurements of the $s$-wave scattering length between interacting atoms \cite{Rey2007}. Here we study the potential application of the model to high-precision force sensing using linear ion crystal. In the following we wish to determine the force amplitude $f_{d,x}$ by measuring the expectation values of the collective spin operator $\hat{J}_{z}$. For this goal let us assume that the system is prepared in the product state $\hat{\rho}(0)=\hat{\rho}_{\rm spin}\otimes\hat{\rho}_{\rm osc}$ where $\hat{\rho}_{\rm osc}$ is the density operator for the quantum oscillator and $\hat{\rho}_{\rm spin}=|\Psi(0)\rangle\langle\Psi(0)|$ with $|\Psi(0)\rangle=\sum_{m=-j}^{j}d_{m}|j,m\rangle_{x}$ being the initial spin state where $d_{m}$ is the reduced Wigner rotation matrix
\begin{equation}
d_{m}=\sqrt{\frac{(2j)!}{(j+m)!(j-m)!}}[\cos(\theta/2)]^{j+m}[\sin(\theta/2)]^{j-m}.
\end{equation}
According to the effective model (\ref{OATM}) the expectation value of $\hat{J}_{z}$ evolves in time as
\begin{equation}
\langle \hat{J}_{z}(t)\rangle=j\sin(\theta)(1-\sin^{2}(\theta)\sin^{2}(\xi))^{2j-1}\cos[\varphi_{f}+(2j-1)\kappa],\label{signal}
\end{equation}
where we define $\xi=\chi^{2}t$, $\varphi_{f}=\Omega_{f}t$ and $\kappa=\tan^{-1}(\tan(\xi)\cos(\theta))$ \cite{Boixo2008}. Hence, in order to determine the force amplitude $f_{d,x}$ one needs to measure the phase $\varphi_{f}$. In Fig. (\ref{fig2}) we show the signal as a function of time assuming initial thermal phonon state. Remarkably, due to vanishing the residual spin-phonon interaction, $\hat{H}^{\prime}=0$, the force sensing protocol does not dependent on the initial vibrational state of the linear ion crystal. As a result of that the measured signal ia naturally robust with respect to the thermally induced spin dephasing.

The uncertainty in the estimate of $\Omega_{f}$ from the measured signal $\langle \hat{J}_{z}(t)\rangle$ is given by
\begin{equation}
\delta\Omega_{f}=\frac{\langle\Delta^{2} \hat{J}_{z}\rangle^{1/2}}{\frac{\partial \langle \hat{J}_{z}\rangle}{\partial \Omega_{f}}\sqrt{\nu}},
\end{equation}
where $\langle\Delta^{2} \hat{J}_{z}\rangle^{1/2}=\sqrt{\langle \hat{J}_{z}^{2}\rangle-\langle \hat{J}_{z}\rangle^{2}}$ is the variance of the signal and $\nu$ is the number of times the estimation is repeated. Assuming the particular value $\theta=\pi/2$ of the initial spin superposition state we have
\begin{eqnarray}
\langle\Delta^{2} \hat{J}_{z}\rangle&=&\frac{j}{2}+\frac{j(2j-1)}{4}+\frac{j(2j-1)}{4}\cos^{2(j-1)}(2\xi)\cos(2\varphi_{f})\notag\\
&&-j^{2}\cos^{2(2j-1)}(\xi)\cos^{2}(\varphi_{f}).\label{variance}
\end{eqnarray}
Using Eqs. (\ref{signal}) and (\ref{variance}) one can show that the optimal sensitivity is achieved at the points $\chi^{2}t=2k\pi$ with $k$ integer. At these points the force sensitivity scales as $\delta\Omega_{f}=1/\sqrt{T t N}$ where we use that $\nu=T/t$, with $T$ being the total experimental time. This is the standard quantum limit in accuracy for measurement of $\Omega_{f}$ using initial uncorrelated spin states.

The entangled motional states can be used to improve the force estimation accuracy at the Heisenberg limit \cite{Munro2002}. However, the physical implementation of such states is in practice difficult since they are very sensitive to motional heating. On the other hand the entangled spin states can be used to improve the sensitivity of frequency estimation using Ramsey fringe interferometry \cite{Bollinger1996,Meyer2001,Leibfried2004}. Because our technique relies on the mapping the relevant force information into the spin-degree of freedom we may use the spin entanglement to increase the sensitivity of the force detection. Indeed, let us assume that the system is prepared initially in the maximally correlated $N$-particle Greenberger-Horne-Zeilinger (GHZ) spin state \cite{Greenberger1990} $\left|\Psi(0)\right\rangle=(\left|j,j\right\rangle_{x}+\left|j,-j\right\rangle_{x})/\sqrt{2}$. The time evolution of the state according Eq. (\ref{OATM}) will induce a phase shift proportional to the force. Subsequently, the parity operator $\hat{\Pi}_{\rm s}$ (\ref{parity}) is measured \cite{Meyer2001} which yields uncertainty in the force estimation
\begin{equation}
f_{d,x}\sqrt{T}=\frac{\hbar\delta_{x}}{2Ng_{x}r_{0,x}\sqrt{t}}.\label{sensitivity}
\end{equation}
Assume for example GHZ state with six ions \cite{Leibfried2005,Monz2011}, $\delta_{x}=100$ kHz, $g_{x}=5$ kHz, $r_{0,x}=15$ nm and evolution time $t=10$ ms and using Eq. (\ref{sensitivity}) we estimate force sensitivity of order of $0.1$ yN $\rm{Hz}^{-1/2}$.

\section{Strong coupling regime}\label{strongregime}
\begin{figure}
\includegraphics[width=0.45\textwidth]{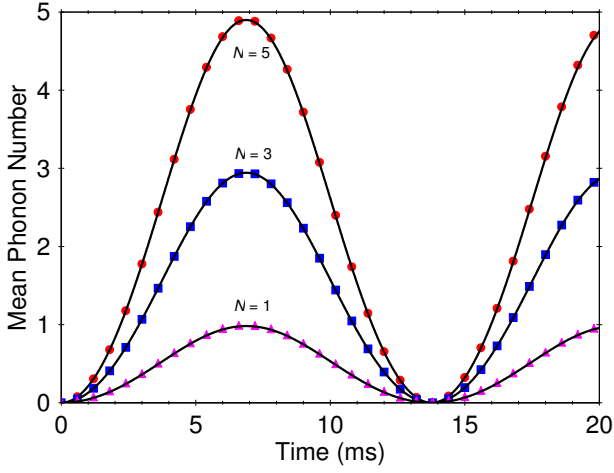}
\caption{(Color online) Time-evolution of the expectation value of $\hat{a}_{x}^{\dag}\hat{a}_{x}$ operator for various number of ions.
We assume an initial $|j,-j\rangle|0_{x}\rangle$. We compare the numerical solution of the time-dependent Schr\"odinger equation with Hamiltonian (\ref{model}) (solid lines) with the solution using effective Hamiltonian (\ref{Hstrong}). The parameters are set to $g_{x}=2.5$ kHz, $\Delta=300$ kHz, $\delta_{x}=0.5$ kHz, $r_{0,x}=14.5$ nm, $f_{d,x}=3$ yN.}
\label{fig3}
\end{figure}

Let us now discuss the case in which the spin-phonon coupling $g_{\alpha}$ is higher than the force detuning $\delta_{\alpha}$ ($g_{\alpha}>\delta_{\alpha}$) which benefits the strong phonon excitations. In contrast to the previous force sensing protocol, the force estimation can now be performed by measuring the mean-phonon number \cite{Maiwald2009}. To this end we assume that the spin frequency $\Delta$ is much larger than the spin-phonon couplings ($\Delta\gg g_{\alpha}$). In this limit the spin dynamics become frozen, thus it can be decoupled from the phonon degree of freedom by using canonical transformation $\hat{U}=e^{\hat{S}}$ with
\begin{equation}
\hat{S}=\frac{2ig_{y}}{\sqrt{N}\Delta}\hat{J}_{x}(\hat{a}_{y}^{\dag}+\hat{a}_{y})-\frac{2i g_{x}}{\sqrt{N}\Delta}\hat{J}_{y}(\hat{a}_{x}^{\dag}+\hat{a}_{x}).
\end{equation}
The resulting effective Hamiltonian becomes
\begin{eqnarray}
&&\hat{H}_{\rm eff}=\hat{H}_{0}+\hat{H}_{\rm ph}+\hat{H}_{F}+\hat{H}^{\prime},\notag\\
&&\hat{H}_{\rm ph}=\frac{2\hbar g_{x}^{2}}{N\Delta}\hat{J}_{z}(\hat{a}_{x}^{\dag}+\hat{a}_{x})^{2}+\frac{2\hbar g_{y}^{2}}{N\Delta}\hat{J}_{z}(\hat{a}_{y}^{\dag}+\hat{a}_{y})^{2},\label{Hstrong}
\end{eqnarray}
where $\hat{H}^{\prime}$ contains high-order terms which can be neglected as long as $\Delta\gg g_{\alpha}$. The Hamiltonian (\ref{Hstrong}) is diagonal in the collective spin basis, thereby the spin-degree of freedom can be traced out giving $N+1$ orthogonal sub-spaces corresponding to each of the collective spin states $\left|j,m\right\rangle$. In addition to it, the Hamiltonian (\ref{Hstrong}) is quadratic in the bosonic operators, thus it can be analytically diagonalized. Let us assume that the system is initially prepared in the spin state $\left|j,-j\right\rangle$. The corresponding bosonic Hamiltonian becomes
\begin{eqnarray}
&&\hat{H}_{\rm eff}=\hat{H}_{x,\rm ph}+\hat{H}_{y,\rm ph}\notag\\
&&\hat{H}_{\alpha,\rm ph}=\hbar\delta_{\alpha}\hat{a}_{\alpha}^{\dag}\hat{a}_{\alpha}-\frac{\hbar g_{\alpha}^{2}}{\Delta}(\hat{a}_{\alpha}^{\dag}+\hat{a}_{\alpha})^{2}
+\frac{f_{d,\alpha}r_{0,\alpha}}{2}(\hat{a}_{\alpha}^{\dag}+\hat{a}_{\alpha}),\label{Hboson}
\end{eqnarray}
\begin{figure}
\includegraphics[width=0.45\textwidth]{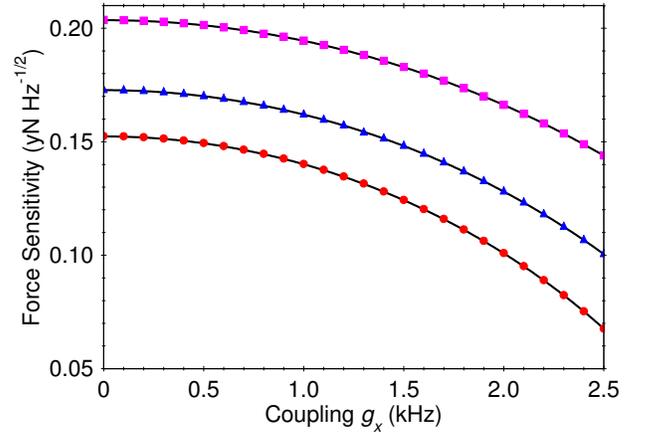}
\caption{(Color online) Force sensitivity as a function of the coupling $g_{x}$ for a single trapped ion. We compare the exact numerical solution for $\delta_{x}=0.14$ kHz (dots), $\delta_{x}=0.18$ kHz (triangles) and $\delta_{x}=0.25$ kHz (squares) with the analytical expression (\ref{fmin}) (solid line). The other parameters are set to $\Delta=270$ kHz and $r_{0,x}=14.5$ nm.  }
\label{fig4}
\end{figure}
Note that the model (\ref{Hboson}) has been studied in the context of a quantum phase transition \cite{Bakemeier2012,Hwang2015} without the force symmetry breaking term $\hat{H}_{F}$ . The unitary propagator corresponding to the Hamiltonian (\ref{Hboson}) can be written as $\hat{U}(t)=\hat{U}_{x}(t)\hat{U}_{y}(t)$, where
\begin{equation}
\hat{U}_{\alpha}(t)=\hat{D}^{\dag}(\epsilon_{\alpha})\hat{S}^{\dag}(\nu_{\alpha})e^{-i\upsilon_{\alpha} t \hat{a}_{\alpha}^{\dag}\hat{a}_{\alpha}}\hat{S}(\nu_{\alpha})\hat{D}(\epsilon_{\alpha}).
\end{equation}
Here $\upsilon_{\alpha}=\delta_{\alpha}\sqrt{1-\lambda^{2}_{\alpha}}$ with $\lambda^{2}_{\alpha}=4g_{\alpha}^{2}/\delta_{\alpha}\Delta$. $\hat{D}(\epsilon_{\alpha})=e^{\epsilon_{\alpha}(\hat{a}_{\alpha}^{\dag}-\hat{a}_{\alpha})}$ is a displacement operator with amplitude $\epsilon_{\alpha}=\sqrt{N}f_{d,\alpha}r_{0,\alpha}/\delta_{\alpha}(1-\lambda_{\alpha}^{2})$ that is proportional to the external force, and respectively $\hat{S}(\nu_{\alpha})=e^{\nu_{\alpha}(\hat{a}^{\dag2}_{\alpha}-\hat{a}_{\alpha}^{\dag})}$ is the squeeze operator with squeezing parameter $\nu_{\alpha}=-\frac{1}{4}\ln(1-\lambda_{\alpha}^{2})$.

In Fig. (\ref{fig3}) we show the time-evolution of $\langle \hat{a}_{x}^{\dag}\hat{a}_{x}\rangle$.  The corresponding signal-to-noise ratio ${\rm SNR}=\langle \hat{a}^{\dag}_{\alpha}\hat{a}_{\alpha}\rangle/\langle\Delta^{2} \hat{a}^{\dag}_{\alpha}\hat{a}_{\alpha}\rangle^{1/2}$ equals to one determines the minimal detectable force. We find that the optimal sensitivity is achieved at the points $\upsilon_{\alpha}t=k\pi$ with $k$ odd number. At these points the signal becomes $\langle \hat{a}^{\dag}_{\alpha}a_{\alpha}\rangle=4\epsilon_{\alpha}^{2}$ with variance of the signal $\langle\Delta^{2} \hat{a}_{\alpha}\hat{a}_{\alpha}\rangle^{1/2}=2\epsilon_{\alpha}$. The minimal detectable force is given by
\begin{equation}
f_{d,\alpha}^{\rm min}=\frac{\hbar\pi\sqrt{1-\lambda_{\alpha}^{2}}}{tr_{0,\alpha}\sqrt{N}}.\label{fmin}
\end{equation}
The result (\ref{fmin}) indicates that for a given force detuning $\delta_{\alpha}$ one can improve the respective force sensitivity limit by increasing the coupling $g_{\alpha}$ and thus $\lambda_{\alpha}$, while keeping the constrain $\Delta\gg g_{\alpha}$, see Fig. \ref{fig4}. Note that here we focus on the case $\lambda_{\alpha}\leq 1$. On the other hand, $\lambda_{\alpha}>1$ leads to high phonon generation which however could break the Lamb-Dicke regime \cite{Wineland1998}. Additionally, we find no major difference in the force sensitivity from the example with $\lambda_{\alpha}\leq 1$.

Let us compare our minimal detectable force assuming single trapped ion $N=1$, with those using a simple harmonic oscillator as a force sensor. For simplicity we assume that the single ion force sensor is sensitive only to one of the force components, namely $f_{d,x}$ with detuning $\delta_{x}$. In that case the quantum probe sensitive to the $f_{d,x}$ is represented by the Rabi Hamiltonian, so that we set $g_{y}=0$ in Eq. (\ref{Hboson}). The minimal detectable force for the harmonic oscillator force sensor is $f_{\rm HO}^{\rm{min}}=\hbar\pi/tr_{0,x}$ which is achieved at $\delta_{x}t=k\pi$ with $k$ odd integer. Comparing with (\ref{fmin}) we conclude that $f_{d,x}^{\rm min}<f_{\rm HO}^{\rm min}$, which indicates that this limit can be overcome within the evolution time $t=k\pi/\delta_{x}\sqrt{1-\lambda_{x}^{2}}$. Moreover, the best sensitivity is typically achieved when the time-varying force alternate with resonance with the motional frequency of the harmonic oscillator. In that case the minimal detectable force is $f_{\rm HO}^{\rm min}=2\hbar/r_{0,x}t$ \cite{Maiwald2009,Munro2002}. Tuning the ratio $\lambda_{x}$ in Eq. (\ref{fmin}) such that $\lambda_{x}>\sqrt{1-4/\pi^{2}}$ we can overcome this force sensitivity limit. For example, using the parameters in Fig. \ref{fig4} with detuning $\delta_{x}=0.14$ kHz and coupling $g_{x}=2.5$ kHz the corresponding force sensitivity is $68$ xN ($10^{-27}$ N) per $\sqrt{\rm{Hz}}$ which is achieved approximately for evolution time $t=40$ ms, while the force sensitivity for the simple harmonic oscillator at the same evolution time is $74$ xN per $\sqrt{\rm{Hz}}$. However, in order to observe such high force sensitivity the evolution time must be short compared with the decoherence time due to the motional heating. For the latter example, this requires very low heating rate of order of $\langle \dot{n}\rangle=1$ s$^{-1}$ which can be achieved for example in a cryogenic ion trap \cite{Brownnutt2015}.
\begin{figure}
\includegraphics[width=0.45\textwidth]{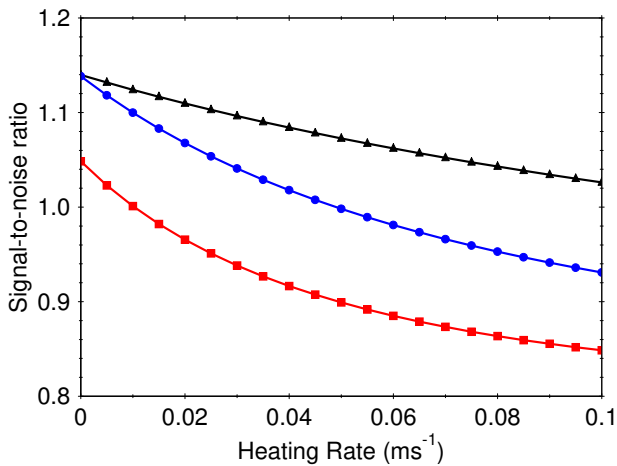}
\caption{(Color online) Signal-to-noise ratio versus heating rate for various $\delta_{x}$. We solve the master equation numerically  (\ref{master}) for single trapped assuming for $g_{y}=0$. We set $g_{x}=25$ kHz, $\Delta=2.7$ MHz, and $f_{d,x}=1.05f_{d,x}^{\rm min}$, $\delta_{x}=1.4$ kHz (squares), $f_{d,x}=1.14f_{d,x}^{\rm min}$, $\delta_{x}=2.1$ kHz (circles) and $\delta_{x}=3.0$ kHz (triangles).}
\label{fig5}
\end{figure}

We examine the dependence on the motional heating of our force sensing protocol by numerical integration of the master equation
\begin{eqnarray}
\frac{d\hat{\rho}}{dt}&=&-\frac{i}{\hbar}[\hat{H},\hat{\rho}]+\frac{\gamma_{\rm dec}}{2}(\bar{n}+1)(2\hat{a}\hat{\rho} \hat{a}^{\dag}-\hat{a}^{\dag}\hat{a}\hat{\rho}-\hat{\rho}\hat{a}^{\dag}\hat{a})\notag\\
&&+\frac{\gamma_{\rm dec}}{2}\bar{n}(2\hat{a}^{\dag}\hat{\rho}\hat{a}-\hat{a}\hat{a}^{\dag}\hat{\rho}-\hat{\rho}\hat{a}\hat{a}^{\dag}),\label{master}
\end{eqnarray}
where $\gamma$ is the system decay rate and $\bar{n}$ is the mean number of quanta in the reservoir. In the limit $\bar{n}\gg 1$ the system is characterized by decoherence time $t_{\rm dec}=1/\bar{n}\gamma$ and the motional heating rate is $\langle \dot{n}\rangle=1/t_{\rm dec}$. Figure \ref{fig5} shows the signal-to-noise ratio as a function of the heating rate for various detuning $\delta_{x}$. As evident, the effect of $\langle\dot{n}\rangle$ on the SNR is weaker for higher $\delta_{x}$, because in that case the relevant phonon degree of freedom becomes less excited. For example, we estimate force sensitivity approximately to $0.4$ yN/$\sqrt{\rm Hz}$ assuming $\langle\dot{n}\rangle=0.05$ $\rm{ms}^{-1}$, $\delta_{x}=2.1$ kHz, $g_{x}=25$ kHz, $\Delta=2.7$ MHz within evolution time $t=2$ ms.

\section{Conclusions}\label{conclusions}
In conclusion, we have introduced sensing protocols capable to measure amplitude of the time-varying forces that are off-resonance with the trap frequencies of the ion chain. Using quantum probe described by the Dicke model, far-detuned forces with detuning much higher than the spin-boson coupling can be efficiently measured by mapping the relevant force information into the collective spin-degree of freedom. Thanks to that we have shown that the force sensitivity can be improved by using initial spin correlated states, leading to Heisenberg limited sensitivity. We have shown that the proposed force sensing protocol is robust with respect to the thermally induced dephasing, which prolong the coherence time and thus improves the force sensitivity. We have also considered sensing protocol capable to detect forces with detuning smaller that the spin-boson coupling. In that case, the relevant force information can be extracted by measuring the mean-phonon number. We have shown that thanks to the strong spin-phonon coupling in the quantum Rabi model the force sensitivity could overcome those using a simple harmonic oscillator as a force sensor.

\end{document}